\documentclass[12pt]{article}

\usepackage{graphicx}
\usepackage{epsfig}
\usepackage{color}
\include{psfig}

\begin{document}
\newcommand{\todo}[1]{{\em \small {#1}}\marginpar{$\Longleftarrow$}}   
\newcommand{\labell}[1]{\label{#1}\qquad_{#1}} 
\newcommand{\ud}{\mathrm{d}}

\rightline{DCPT-04/29}   
\rightline{hep-th/0409188}   
\vskip 1cm


\begin{center} 
{\Large \bf On uniqueness of charged Kerr-AdS black holes in five dimensions}
\end{center} 
\vskip 1cm   
  
\renewcommand{\thefootnote}{\fnsymbol{footnote}}   
\centerline{\bf   
Owen Madden\footnote{O.F.Madden@durham.ac.uk} and Simon 
F. Ross\footnote{S.F.Ross@durham.ac.uk}}    
\vskip .5cm   
\centerline{ \it Centre for Particle Theory, Department of  
Mathematical Sciences}   
\centerline{\it University of Durham, South Road, Durham DH1 3LE, U.K.}   
  
\setcounter{footnote}{0}   
\renewcommand{\thefootnote}{\arabic{footnote}}


\begin{abstract}  
We show that the solutions describing charged rotating black holes in
five-dimensional gauged supergravities found recently by Cveti\v c,
L\"u and Pope \cite{clp1,clp2} are completely specified by the mass,
charges and angular momentum. The additional parameter appearing in
these solutions is removed by a coordinate transformation and
redefinition of parameters. Thus, the apparent hair in these solutions
is unphysical.
\end{abstract}

\section{Introduction}

The question of black hole uniqueness in higher dimensions has been
attracting considerable attention since the discovery by Emparan and
Reall~\cite{ring} of an asymptotically flat black ring in five
dimensions. This is a solution with a regular event horizon of
topology $S^2 \times S^1$, supported against collapse by angular
momentum. Since there is a rotating black hole which carries the same
mass and angular momentum as this solution, this represents a
breakdown of the usual no-hair behaviour: the solution is not uniquely
determined by the asymptotic conserved charges. Studies of charged
rings have uncovered further examples of
non-uniqueness~\cite{cring1,infinite,cring2,conc}.

So far, all these examples involve discrete forms of non-uniqueness:
the existence of some finite number of solutions with the same
asymptotic charges. Although the classical solutions
in~\cite{infinite} involved a continuous parameter, this
parameter can be physically interpreted in terms of a local charge
carried by the ring, so it will be quantized in the fundamental
theory. Furthermore, bounds on this charge for the existence of a
regular event horizon imply that these solutions give a finite number
of solutions with given energy. This is consistent with the
expectation that the fundamental quantum theory has a finite number of
states of given energy in finite volume. We would expect that in
general distinct classical solutions must correspond to different
quantum states. 

A new example with an apparent continuous non-uniqueness was recently
found in~\cite{clp1,clp2}. They constructed solutions describing
rotating charged black holes in five-dimensional gauged supergravity,
with the two angular momentum parameters set equal. Studying first a
case with a single $U(1)$ gauge field~\cite{clp1}, they found
solutions with four parameters: the mass, charge, angular momentum,
and one additional parameter. In~\cite{clp2}, they extended this to a
$U(1)^3$ theory with independent charges and found a solution
depending on six parameters. The solutions thus appeared to involve a
continuous non-uniqueness, which initially appeared to have physical
consequences. In light of the previous discussion, we want to
understand the physical significance of this extra parameter, to see
how the apparent contradiction with our expectation that there should
be a finite number of solutions of given energy is resolved. In
addition, these solutions appear to provide a first example of
non-uniqueness involving black holes with a spherical horizon
topology, so it would clearly be interesting to understand them in
more detail.

In this paper, we will show that the extra parameter in the solutions
of~\cite{clp1,clp2} is unphysical, representing a purely coordinate
degree of freedom. In~\cite{clp1}, it was observed that if the charge
$Q$ vanished, the additional parameter could be removed by a
coordinate transformation and redefinition of the other parameters. In
the next section, we extend this to the case with $Q \neq 0$. We
discuss the extremal limits in terms of these parameters, showing how
the solutions of~\cite{klemms,gutreall} are recovered in our
parametrization, and briefly discuss the reduction to other known
solutions. We also discuss the possibility of discrete non-uniqueness,
and show that although the relation between our parametrization of the
solutions and the physical mass and charges is non-linear, there is
only one black hole solution for given mass, angular momentum and
charges. We present a similar argument for the $U(1)^3$ solutions
of~\cite{clp2} in section~\ref{3ch}.

Thus, these solutions do not in fact present new examples of
non-uniqueness. However, as stressed in~\cite{clp1,clp2}, they do
provide interesting testing grounds for the AdS/CFT correspondence,
and generalise known solutions in interesting ways. (In particular,
they provide non-extreme versions of the interesting supersymmetric
asymptotically AdS solution found in~\cite{gutreall}.) 

\section{Charged Kerr-de Sitter Black Holes in five dimensions}
\label{1ch}

We consider first the solution obtained in~\cite{clp1}, describing a
charged rotating black hole with a cosmological
constant. The solution can be written in the simplest form by
introducing the left-invariant one-forms $\sigma_i$ on $S^3$,
\begin{eqnarray}
\sigma_1 &=& \cos \psi d \theta + \sin \psi \sin \theta d\phi, \\
\sigma_2 &=& -\sin \psi d \theta + \cos \psi \sin \theta d\phi, \\
\sigma_3 &=& d\psi + \cos \theta d\phi.
\end{eqnarray}
The solution then takes the form 
\begin{eqnarray} 
\mathrm{d}s^2&=&-\frac{r^2 W}{4b^2}\mathrm{d}t^2+\frac{1}{W}\mathrm{d}r^2
+\frac{r^2}{4}(\sigma_1^2+\sigma_2^2)+b^2(\sigma_{3}+f\mathrm{d}t)^2,
\\ A&=&\frac{\sqrt{3}Q}{r^2}(\mathrm{d}t-\frac{1}{2}J\sigma_{3}),
\end{eqnarray}
where
\begin{eqnarray} 
b^2&=&\frac{r^2}{4}\left[1-\frac{J^2
Q^2}{r^6}+\frac{2J^2(M+Q)}{r^4}\right], \\
f&=&-\frac{J}{2b^2}\Big(\lambda\beta
r^2+\frac{2M+Q}{r^2}-\frac{Q^2}{r^4}\Big),\\ W&=&1-\lambda
r^2-\frac{1}{r^2}\left[2\lambda J^2(M+Q)+2(1-\lambda\beta
J^2)^2(M+Q)-2Q(1-\lambda\beta J^2)\right] \nonumber \\ &&
+\frac{1}{r^4}\left\{(1-\lambda\beta J^2)^2Q^2+J^2[\lambda
Q^2+2(M+Q)]\right\} .
\end{eqnarray} 
This is a solution of minimal gauged five-dimensional supergravity.
It appears to depend on four parameters, $(M,J,Q,\beta)$.  The first
three can be related to the mass, angular momentum and charge, but the
physical interpretation of the fourth is obscure: we will show that
this solution can in fact be written in terms of only three
parameters.

The first step is to transform to a frame in which the metric is
non-rotating at infinity, so that it approaches the usual diagonal
form of the AdS metric at large distances, by making the shift 
\begin{equation}
\tilde{\sigma}_3=\sigma_{3}+2\lambda\beta J\mathrm{d}t. 
\end{equation}
We then have
\begin{eqnarray} 
\mathrm{d}s^2&=&-\frac{r^2
W}{4b^2}\mathrm{d}t^2+\frac{1}{W}\mathrm{d}r^2 +\frac{r^2}{4}
(\sigma_1^2+\sigma_2^2)+b^2(\tilde{\sigma}_{3}+\tilde{f}\mathrm{d}t)^2,
\\ A&=&\frac{\sqrt{3}Q}{r^2}\left[(1-\lambda\beta
J^2)\mathrm{d}t-\frac{1}{2}J\tilde\sigma_{3}\right],
\end{eqnarray}
where
\begin{equation}
\tilde{f}=-\frac{J}{2b^2}\left[\frac{2(M+Q)(1-\lambda\beta
  J^2)-Q}{r^2}-\frac{Q^2(1-\lambda\beta J^2)}{r^4}\right]. 
\end{equation}
It is convenient to exchange $\beta$ for a new parameter $\delta$ (as
in~\cite{hsy}) 
\begin{equation}
\delta=(1-\lambda\beta J^2).
\end{equation}
We then see that the gauge field becomes 
\begin{equation}
A = \frac{\sqrt{3} \delta Q}{r^2} dt - \frac{\sqrt{3}QJ}{2 r^2}
\tilde\sigma_3,  
\end{equation}
so the physical gauge charge is related to $\delta Q$. We also see
that $\tilde f$ involves the parameters only in the combinations $J
\delta (M+Q)$, $\delta Q$ and $JQ$.  This suggests that we define
the following new  parameters: 
\begin{eqnarray}
q&=&Q\delta, \\ \nonumber
p&=&\delta^2(M+Q), \\
j&=&J\over\delta.
\end{eqnarray}
We then find that the solution depends only on the three parameters
$(q,j,p)$, and has no remaining dependence on $\delta$: 
\begin{eqnarray} \label{metric}
\mathrm{d}s^2&=&-\frac{r^2
W}{4b^2}\mathrm{d}t^2+\frac{1}{W}\mathrm{d}r^2 +\frac{r^2}{4}
(\sigma_1^2+\sigma_2^2)+b^2(\tilde{\sigma}_{3}+\tilde{f}\mathrm{d}t)^2,
\\ A&=&\frac{\sqrt{3} q}{r^2} dt - \frac{\sqrt{3}qj}{2 r^2}
\tilde\sigma_3,  
\end{eqnarray}
where
\begin{eqnarray}
b^2&=&\frac{r^2}{4}\Big(1-\frac{j^2 q^2}{r^6}+\frac{2j^2 p}{r^4}\Big), \\
\tilde{f}&=&-\frac{j}{2b^2}\Big(\frac{2p-q}{r^2}-\frac{q^2}{r^4}\Big), \\
W&=&1-\lambda r^2-\frac{1}{r^2}(2\lambda j^2 p+2p-2q)+\frac{1}{r^4}\big[q^2+j^2(\lambda q^2+2p)\big].
\end{eqnarray}
It is perhaps worth noting that this expression for $W$ can be
rewritten as
\begin{equation}
W = 1 -4 \lambda b^2 - \frac{1}{r^2}(2p-2q) + \frac{1}{r^4}
(q^2+2pj^2).
\end{equation}

Thus, we see that the solution actually only depends on three
parameters, corresponding to the three conserved quantities
$(M,Q,J)$. The relation of two of the parameters to these conserved
quantities is fairly direct: from the form of the gauge potential, we
see that 
\begin{equation}
Q=q.
\end{equation}
The angular momentum $J$ can be calculated using the Komar integral technique
and is given by
\begin{equation} \label{angmom}
J=\frac{\pi}{4}j(2p-q),
\end{equation}
so $j$ is analogous to the angular momentum per unit
mass parameter $a$ in the usual Kerr solution. 

The remaining parameter $p$ is related to the freedom to specify the
mass, but in a less simple way. We can define a thermodynamic mass as
in~\cite{gpp} by insisting that it satisfy the first law of
thermodynamics for charged rotating black holes,
\begin{equation}
\ud M=T\ud S+2\Omega_H\ud J +\Phi_H\ud Q,
\end{equation}
where $\Phi_H$ is the co-rotating electric potential evaluated on the
horizon. This gives the mass 
\begin{equation} \label{mass}
M=\frac{\pi}{4}(3p-3q-\lambda p j^2).
\end{equation}
Note that although this looks like a linear relation, if we solve
(\ref{angmom}) for $j$ in terms of $J$, (\ref{mass}) will become a
cubic equation for $p$ in terms of $M$.

The reduction of this solution to previously known metrics was
discussed at length in~\cite{clp1}. We will briefly revisit this issue
to illuminate our parametrization of the solution. Considering first
the known BPS solutions, we note that the form above reduces to the
solution in ~\cite{klemms} if we set $p=0$ (after a redefinition of
the radial coordinate, $r^2 \to r^2 - q$), so this choice of
parameters is well-adapted to this limit, while recovering the
solution of~\cite{gutreall} requires a more complicated choice: To
recover their solution, we write $\lambda = -1/\ell^2$ and set
\begin{eqnarray}
q &=& \left( 1 + \frac{R_0^2}{2\ell^2} \right) R_0^2, \\
p &=&2 \left(1 + \frac{R_0^2}{2 \ell^2} \right)^2 R_0^2, \\
j &=& \frac{ \epsilon l R_0^2}{2} \left( 1 + \frac{R_0^2}{2 \ell^2}
\right)^{-1}.
\end{eqnarray} 
From the results of~\cite{gutreall}, we see that 
for this choice of parameters, the solution has a degenerate horizon
(a double root of $W=0$) at $r=R_0$. The condition that $W=0$ have a
double root at $r=R_0$ in general implies
\begin{equation}
2\lambda j^2 p + 2p -2q = 2R_0^2 -3\lambda R_0^4
\end{equation}
and
\begin{equation}
q^2 + j^2(\lambda q^2 + 2p) = R_0^4 - 2 \lambda R_0^2.
\end{equation}
Since these are two conditions on the three parameters $(q,p,j)$,
there is a family of solutions with degenerate horizons with one 
extra parameter, generalising the solution found
in~\cite{gutreall}. However, from the analysis in~\cite{gutreall}, we
expect that only the solution found there is BPS. 

The other simple special cases are when $j=0, q=0$ or
$\lambda=0$. When $j=0$, if we set $p = m + q$, the solution reduces
to the RNAdS black hole studied in~\cite{cejm}. When $q=0$, if
we set
\begin{equation}
\lambda = -l^2, \quad j = a, \quad p = \frac{m}{(1-a^2 l^2)^3}
\end{equation}
the solution reduces to a special case of the five-dimensional Kerr-AdS
metric obtained in~\cite{hht}, where the two angular momentum
parameters are equal, $a=b$. Relating the form of the metric used
in~\cite{hht} to (\ref{metric}) requires a shift of the angular
coordinate to make the metric in~\cite{hht} asymptotically diagonal
and a redefinition of the radial coordinate, $r^2 \to r^2 + a^2$.
When $\lambda=0$, the solution is related to a special case of the
general solution obtained in~\cite{cy} where the two angular momentum
parameters are equal, $l_1 = l_2 = l$, and the three charge parameters
are equal, $\delta_e = \delta_{e1} = \delta_{e2}$. The relation
between the parameters is
\begin{equation}
q = m \sinh 2\delta_e, \quad j = l e^{-\delta_e}, \quad p = m e^{2\delta_e}.
\end{equation}
Relating the metric in~\cite{cy} to (\ref{metric}) again requires a
redefinition of the radial coordinate, $r^2 \to r^2 +l^2 + 2m \sinh^2
\delta_e$.  

As discussed in~\cite{clp1}, the physically interesting solutions are
those with a regular event horizon and no closed timelike curves
outside the horizon. That is, we want to consider parameter values for
which there is some $r_+$ such that  $W(r_+)=0$, $W'(r_+) \geq 0$ (so that
$r_+$ is the outer event horizon) and $b^2(r_+) >0$. Written
explicitly in terms of our parameters, these conditions are
\begin{equation} \label{rplus}
r_+^4 (1- 4 \lambda b_+^2) - 2 r_+^2(p-q)+ q^2 +
2pj^2=0,
\end{equation}
where we have introduced the notation $b_+^2 \equiv b(r_+)^2$, 
\begin{equation}
-\lambda r_+^6 + 2r_+^2(\lambda j^2p+p-q) -2
 [q^2 + j^2(\lambda q^2 + 2p)] \geq 0 ,
\end{equation}
and
\begin{equation}
r_+^6 + 2 j^2 p r_+^2  - j^2 q^2 > 0.
\end{equation}
These conditions will impose some constraints on the values of
$(p,j,q)$. For example, the requirement that (\ref{rplus}) have a real
positive root for $r_+^2$ implies that 
\begin{equation}
(p-q)^2  \geq (q^2+2p j^2)(1 - \lambda b_+^2).
\end{equation}
Unfortunately, although our coordinate transformation and redefinition
of parameters simplifies the functions somewhat, it is still difficult
to analyse the full set of constraints. 

An alternative approach is to use the above relations to replace the
parameters $j$ and $p$ in the metric by $r_+$ and $b_+^2$, thereby
automatically incorporating the constraint $b^2(r_+) >0$ for the
absence of naked closed timelike curves. We can easily determine the
relation between $j$ and $b_+$,
\begin{equation}
j^2 = \frac{r_+^4 (4b_+^2 - r_+^2)}{(2pr_+^2-q^2)}
\end{equation}
and substituting this into (\ref{rplus}) then gives a quadratic equation
to solve for $p$, determining it in terms of $r_+, b_+$ and $q$. The
resulting form of the metric is, unfortunately, rather messy and
unenlightening, and it would still be necessary to somehow incorporate
the constraint that $W'(r_+) \geq 0$, which will restrict the possible
values of $r_+$ and $b_+$ for given $q$. 

So far we have discussed continuous non-uniqueness. We should also
consider the possibility of discrete non-uniqueness\footnote{We thank
  the referee for raising this issue.}. As noted
previously, (\ref{mass}) gives a cubic equation for $p$ as a function
of the physical parameters $M, J, Q$, so it is possible that we will
have more than one black hole solution for a given mass and charge. 
If we define the dimensionless quantities 
\begin{eqnarray}
\gamma&=&\lambda(2p-q),~\tilde M=\frac{4\lambda}{\pi}M, \nonumber \\
\tilde J&=&\frac{4\lambda^2}{\pi}J,~\tilde Q=\lambda Q
\end{eqnarray}
then (\ref{angmom}) tells us 
\begin{equation}
j = \frac{\tilde{J}}{\lambda \gamma},
\end{equation}
and substituting this into (\ref{mass}) gives 
\begin{equation}
\gamma^3-(2\tilde M+3\tilde Q)\gamma^2+\tilde J^2\gamma+\tilde J^2 \tilde Q=0.
\label{gam}
\end{equation}
This is a cubic equation in $\gamma$ so for each $(M,J,Q)$ there are
 possibly three different solutions. However, for a solution to
 correspond to a black hole, $W(r_+)=0$ must have at least one
 positive real root, at which $b^2(r_+) >0$. We performed a numerical
 analysis to check whether there are values of $(M,J,Q)$ for which
 more than one of the roots of ~(\ref{gam}) satisfy these conditions.
 We found that at most one root of~(\ref{gam}) has a black hole
 interpretation, thus ruling out any discrete non-uniqueness. A
 representative plot is shown in figure 1.

\begin{figure}[h]
\centering
\input{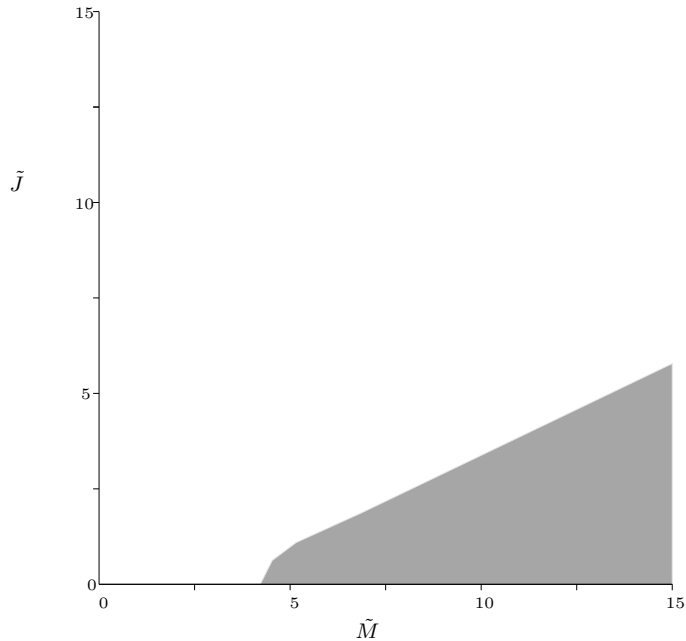}
\caption{Plot showing the parameters for which we have one black hole
  solution (shaded region) and for which we have no solutions with
  horizons (white region) for $\tilde Q=+1$.}
\label{pic2}
\end{figure}

\section{$U(1)^3$ case}
\label{3ch}

In a further paper~\cite{clp2}, the previous solution was generalised
to a class of non-extremal charged rotating black hole solutions in
the five dimensional $U(1)^3$ gauged theory of $\mathcal{N}=2$
supergravity coupled to two vector multiplets. This theory now has
three gauge fields $A^i$, $i =1,2,3$, and two scalars $\varphi_1,
\varphi_2$. The solutions they obtain can be written as
\begin{eqnarray} 
\mathrm{d}s^2&=&-\frac{RY}{f_{1}}\mathrm{d}t^2+\frac{Rr^2}{Y}\mathrm{d}r^2
+\frac{1}{4}R(\sigma_1^2+\sigma_2^2)+\frac{f_{1}}{4R^2}(\sigma_{3}-2\frac{f_2}{f_1}\mathrm{d}t)^2
\\ A^i&=&\frac{\mu}{r^2 H_i}\left[s_i c_i\mathrm{d}t-\frac{1}{2}l(c_i
s_j s_k-s_i c_j c_k)\sigma_{3}\right], \\ e^{\frac{2}{\sqrt{6}}
  \varphi_1}  &=& X_3, \quad e^{\sqrt{2} \varphi_2} =
\frac{X_2}{X_1},
\end{eqnarray}
where 
\begin{equation}
X_i=\frac{R}{r^2 H_i}, \quad i=1,2,3
\end{equation}
\begin{equation}
R=r^2 \Big(\prod_{i}^3 H_{i}\Big)^\frac{1}{3},  \quad
H_{i}=1+\frac{\mu s_{i}^2}{r^2}, 
\end{equation}
 $s_i$ and $c_i$ are shorthand for 
\begin{equation}
s_i\equiv \sinh\delta_i,~~c_i\equiv \cosh\delta_i,
\end{equation}
and
\begin{eqnarray} 
f_1&=&R^3+\mu l^2 r^2+\mu^2 l^2
\left[2(\prod_{i}c_{i}-\prod_{i}s_{i})\prod_{j}s_{j}-\sum_{i<j}s_i^2
s_j^2 \right], \nonumber\\
f_2&=&\gamma l\lambda R^3+\mu l(\prod_{i}c_{i}-\prod_{i}s_{i})r^2+\mu^2 l\prod_i s_i, \\
f_3&=&\gamma^2 l^2\lambda^2 R^3+\mu l^2\lambda\left[2\gamma(\prod_{i}c_{i}-\prod_{i}s_{i})-1-\gamma^2 l^2\lambda\right]r^2  +\mu l^2 \\ && -\lambda(1+\gamma^2 l^2\lambda)\mu^2 l^2\left[2(\prod_{i}c_{i}-\prod_{i}s_{i})\prod_{j}s_{j}-\sum_{i<j}s_i s_j\right]+2\lambda\gamma\mu^2 l^2\prod_i s_i, \nonumber \\
Y&=&f_3-\lambda(1+\gamma^2 l^2\lambda)R^3+r^4-\mu r^2.
\end{eqnarray}

This solution seems to depend on six non-trivial parameters
$(\mu,\delta_1,\delta_2,\delta_3,l,\gamma)$. Our aim is to show that
this depends on only five independent parameters. Again we begin by
moving to coordinates in which the metric is asymptotically diagonal by
setting 
\begin{equation} 
 \tilde{\sigma}_3=\sigma_{3}+2\gamma l\lambda\mathrm{d}t. 
\end{equation}
We then have 
\begin{eqnarray}
\mathrm{d}s^2&=&-\frac{RY}{f_{1}}\mathrm{d}t^2+\frac{Rr^2}{Y}\mathrm{d}r^2
+\frac{1}{4}R(\sigma_1^2+\sigma_2^2)+\frac{f_{1}}{4R^2}(\tilde{\sigma}_3-2\frac{\tilde{f_2}}{f_1}\mathrm{d}t)^2, \\
A^i&=&\frac{\mu}{r^2 H_i}\left\{[s_i c_i+\gamma\lambda l^2(c_i s_j s_k-s_i c_j c_k)]\mathrm{d}t+\frac{1}{2}l(c_i s_j s_k-s_i c_j c_k)\tilde\sigma_{3}\right\},
\end{eqnarray}
where
\begin{eqnarray}
\tilde{f_2} &=& f_{2}-\gamma l\lambda f_1 \\ &=& \left[ \mu
  l(\prod_{i}c_{i}-\prod_{i}s_{i}) - \gamma \mu l^3 \lambda \right]
  r^2 + \mu^2 l\prod_i s_i \nonumber \\ &&- \gamma l \lambda \mu^2 l^2
  \left[2(\prod_{i}c_{i}-\prod_{i}s_{i})\prod_{j}s_{j}-\sum_{i<j}s_i^2
  s_j^2 \right]. \nonumber
\end{eqnarray}
The radial coordinate used here is different from that used in the
previous case: the singularity in this metric will occur at $r^2 =
-\mu s_i^2$, where $\delta_i$ is the smallest of the charge parameters, and not
at $r=0$ as before. We are therefore motivated to make a change of
radial coordinate to a new radial coordinate $\rho$,
\begin{equation}
r^{2}=\rho^2-\frac{1}{3}\sum_{i}\mu s_{i}^2 .
\end{equation}
With this new choice of radial coordinate, 
\begin{equation}
X_{i}=\frac{R}{\rho^2 \tilde{H}_i},
\end{equation}
where
\begin{equation}
R = \rho^2 \Big(\prod_{i}^3
\tilde{H}_{i}\Big)^\frac{1}{3}, \quad
\tilde{H}_{i}=1+\frac{\mu(s_{i}^2-s_j^2) + \mu (s_i^2-s_k^2)}{3\rho^2}, 
\end{equation}
so the scalar fields, which are determined by the $X_i$, depend on the
parameters only through the combinations $\mu (s_i^2 - s_j^2)$. 

As in the previous case, we will find suitable parameters by examining
the gauge fields and the function $\tilde{f_2}$ appearing in the
asymptotic form of the metric. We are thereby led to define the five
independent parameters
\begin{eqnarray}
L&=&\sqrt{\mu} l, \\
\Gamma&=&\sqrt{\mu} \left[(\prod_{i}c_{i}-\prod_{i}s_{i})-\gamma
  l^2\lambda\right],\\ 
r_i&=&\sqrt{\mu} (c_{i}s_{j}s_{k}-s_{i}c_{j}c_{k}).
\end{eqnarray}
This may seem an awkward definition, but the $r_i$ in particular have
several nice properties:
\begin{equation}
r_i^2 - r_j^2 = \mu (s_i^2 - s_j^2),
\end{equation}
so the combinations $\mu (s_i^2 - s_j^2)$ appearing in the
$\tilde{H}_i$, and hence in the scalar
fields, can be written as $(r_i^2 - r_j^2)$. Also,
\begin{equation}
 \mu \left[s_i c_i+\gamma\lambda l^2(c_i s_j s_k-s_i c_j c_k)
  \right] = r_j r_k - \Gamma r_i,
\end{equation}
so the gauge fields can be written as 
\begin{equation}
A^i = \frac{r_j r_k - \Gamma r_i}{ \rho^2 \tilde H_i} dt  +
\frac{L r_i}{2 \rho^2 \tilde{H}_i} \tilde \sigma_3.
\end{equation}
Finally, the metric in the new coordinates is 
\begin{equation}
\mathrm{d}s^2= -\frac{RY}{f_{1}}\mathrm{d}t^2 +
\frac{R\rho^2}{Y}\mathrm{d}\rho^2 +
\frac{1}{4}R(\sigma_1^2+\sigma_2^2) + \frac{f_{1}}{4R^2}
(\tilde{\sigma}_3-2\frac{\tilde{f_2}}{f_1}\mathrm{d}t)^2,
\end{equation}
and after a certain amount of calculation, it is possible to rewrite
$f_1,f_2$ and $Y$ as
\begin{eqnarray}
f_1&=&R^3+L^2\left( \rho^2-\frac{1}{3}\sum_{i}r_{i}^2 \right), \\
\tilde{f_2}&=&L \left( \Gamma \rho^2-\frac{1}{3}\Gamma\sum_{i}r_{i}^2+
r_1 r_2 r_3 \right), \\
Y&=&-\lambda R^3+\rho^4+\Big(\frac{1}{3}\sum_{i}r_{i}^2-\lambda L^2
-\Gamma^2 \Big)\rho^2+\frac{1}{3}\Gamma^2\sum_{i}r_{i}^2-2\Gamma
  r_1 r_2 r_3\nonumber \\&& +\frac{1}{3}\lambda L^2 \sum_{i}r_{i}^2+L^2+\left[\frac{5}{18}(\sum_{i}r_{i}^2)^2-\frac{1}{2}\sum_{i}r_{i}^4\right].
\end{eqnarray}

The solution therefore depends only on five independent parameters,
and a metric will be uniquely fixed by specifying the mass, angular
momentum and three gauge charges. That is, this case is qualitatively
the same as in the previous section. A useful consistency check is that
when the three charges are equal, this reduces to the previous metric
on setting 
\begin{equation}
L = \sqrt{2p} j, \quad \Gamma = -\frac{1}{\sqrt{2p}}(2p-q), \quad r_i
  = \frac{q}{\sqrt{2p}} .
\end{equation}

We have not considered the relation of these parameters to the
physical mass, angular momentum and charges, so there is still the
possibility of some discrete non-uniqueness in this case, but we doubt
this possibility is realised in practice. 

{\large\bf Acknowledgements}

We thank Mirjam Cveti\v c for pointing out the apparent non-uniqueness
in~\cite{clp1,clp2} to us. The work of SFR is supported by the
EPSRC. The work of OM is supported by the PPARC.

\providecommand{\href}[2]{#2}\begingroup\raggedright\endgroup


\begin{thebibliography}{10}

\bibitem{clp1}
M.~Cveti\v c, H.~L\" u, and C.~N. Pope, ``Charged {Kerr-de S}itter
black holes in 
  five dimensions,''
\href{http://xxx.lanl.gov/abs/hep-th/0406196}{{\tt hep-th/0406196}}.

\bibitem{clp2}
M.~Cveti\v c, H.~L\" u, and C.~N. Pope, ``Charged rotating black holes in five
  dimensional {$U(1)^3$ gauged $N = 2$} supergravity,''
\href{http://xxx.lanl.gov/abs/hep-th/0407058}{{\tt hep-th/0407058}}.

\bibitem{ring}
R.~Emparan and H.~S. Reall, ``A rotating black ring in five dimensions,'' Phys.
  Rev. Lett. {\bf 88} (2002) 101101,
\href{http://xxx.lanl.gov/abs/hep-th/0110260}{{\tt hep-th/0110260}}.

\bibitem{cring1}
H.~Elvang and R.~Emparan, ``Black rings, supertubes, and a stringy resolution
  of black hole non-uniqueness,'' JHEP {\bf 11} (2003) 035,
\href{http://xxx.lanl.gov/abs/hep-th/0310008}{{\tt hep-th/0310008}}.

\bibitem{infinite}
R.~Emparan, ``Rotating circular strings, and infinite non-uniqueness of black
  rings,'' JHEP {\bf 03} (2004) 064,
\href{http://xxx.lanl.gov/abs/hep-th/0402149}{{\tt hep-th/0402149}}.

\bibitem{cring2}
H.~Elvang, R.~Emparan, D.~Mateos, and H.~S. Reall, ``Supersymmetric black rings
  and three-charge supertubes,''
\href{http://xxx.lanl.gov/abs/hep-th/0408120}{{\tt hep-th/0408120}}.

\bibitem{conc}
J.~P. Gauntlett and J.~B. Gutowski, ``Concentric black rings,''
\href{http://xxx.lanl.gov/abs/hep-th/0408010}{{\tt hep-th/0408010}}.

\bibitem{klemms}
D.~Klemm and W.~A. Sabra, ``Charged rotating black holes in 5d
  {Einstein-Maxwell-(A)dS gravity},'' Phys. Lett. {\bf B503} (2001) 147--153,
\href{http://xxx.lanl.gov/abs/hep-th/0010200}{{\tt hep-th/0010200}}.

\bibitem{gutreall}
J.~B. Gutowski and H.~S. Reall, ``Supersymmetric{ AdS$_5$} black holes,'' JHEP
  {\bf 02} (2004) 006,
\href{http://xxx.lanl.gov/abs/hep-th/0401042}{{\tt hep-th/0401042}}.

\bibitem{hsy}
Y.~Hashimoto, M.~Sakaguchi, and Y.~Yasui, ``Sasaki-{E}instein twist of
  {Kerr-AdS} black holes,''
\href{http://xxx.lanl.gov/abs/hep-th/0407114}{{\tt hep-th/0407114}}.

\bibitem{gpp}
G.~W. Gibbons, M.~J. Perry, and C.~N. Pope, ``The first law of thermodynamics
  for {Kerr - anti-de Sitter} black holes,''
\href{http://xxx.lanl.gov/abs/hep-th/0408217}{{\tt hep-th/0408217}}.

\bibitem{cejm}
A.~Chamblin, R.~Emparan, C.~V. Johnson, and R.~C. Myers, ``Charged {AdS} black
  holes and catastrophic holography,'' Phys. Rev. {\bf D60} (1999) 064018,
\href{http://xxx.lanl.gov/abs/hep-th/9902170}{{\tt hep-th/9902170}}.

\bibitem{hht}
S.~W. Hawking, C.~J. Hunter, and M.~M. Taylor-Robinson, ``Rotation and the
  ads/cft correspondence,'' Phys. Rev. {\bf D59} (1999) 064005,
\href{http://xxx.lanl.gov/abs/hep-th/9811056}{{\tt hep-th/9811056}}.

\bibitem{cy}
M.~Cveti\v c and D.~Youm, ``General rotating five dimensional black holes of
  toroidally compactified heterotic string,'' Nucl. Phys. {\bf B476} (1996)
  118--132,
\href{http://xxx.lanl.gov/abs/hep-th/9603100}{{\tt hep-th/9603100}}.

\end{thebibliography}
\end{document}